# Simulated Intervention on Cross-Sectional Nested Data: Development of a Multilevel NIRA Approach

Yiming Wu[1], Fei Wang[2,3*]

1 School of Psychology, South China Normal University, 510631, Guangzhou, China
2 CAS Key Laboratory of Behavioral Science, Institute of Psychology, 100101, Beijing, China
3 Department of Psychology, University of Chinese Academy of Sciences, 100049, Beijing, China

**Abstract**

With the rise of the network perspective, researchers have made numerous important discoveries over the past decade by constructing psychological networks. Unfortunately, most of these networks are based on cross-sectional data, which can only reveal associations between variables but not their directional or causal relationships. Recently, the development of the nodeIdentifyR algorithm (NIRA) technique has provided a promising method for simulating causal processes based on cross-sectional network structures. However, this algorithm is not capable of handling cross-sectional nested data, which greatly limits its applicability. In response to this limitation, the present study proposes a multilevel extension of the NIRA algorithm, referred to as multilevel NIRA. We provide a detailed explanation of the algorithm's core principles and modeling procedures. Finally, we discuss the potential applications and practical implications of this approach, as well as its limitations and directions for future research.

***Keywords***: NIRA; Psychological networks; Multilevel modeling; Cross-sectional nested data; Causal inference

# 1 Introduction

Over the past decade, network analysis has emerged as a pivotal conceptual framework and methodological approach in psychological research, finding widespread application across diverse domains of psychology (Burger et al., 2023). Correspondingly, studies employing network methodologies have attracted considerable attention (Contreras et al., 2019), with the majority focusing on the field of psychopathology (Cramer et al., 2016; Cramer, Waldorp, van der Maas, & Borsboom, 2010). To some extent, network analysis effectively addresses the limitations inherent in traditional approaches to psychopathology. Specifically, conventional models of psychopathology implicitly assume that observed symptoms reflect the influence of latent variables, thereby overlooking the fact that symptoms often exhibit strong interrelations and may not fully capture the complexity of psychological phenomena (Borsboom, 2008). Network theory overcomes this issue by emphasizing that symptoms mutually influence one another, and that both the symptoms themselves and the associations between them constitute the disorder; symptoms are not merely downstream effects of latent constructs (Borsboom, 2017; Borsboom & Cramer, 2013). Guided by this perspective, researchers have employed psychological network models to map the underlying structure of symptom interactions: symptoms are conceptualized as nodes, and the statistical relationships between symptoms as edges connecting those nodes (Epskamp et al., 2012, 2018; Epskamp & Fried, 2018).

From the standpoint of edge directionality, networks can be classified as directed or undirected. Directed networks feature edges with an intrinsic orientation, representing unidirectional effects from node A to node B. In psychological research, examples include cross-lagged panel networks—where a variable at Time 1 influences another at Time 2, thus exhibiting directionality (Bringmann et al., 2016)—and Bayesian networks, which capture hierarchical dependencies by modeling the probability of event B given event A, with A situated hierarchically above B (Kitson et al., 2023). In contrast, undirected networks display edges without inherent direction, indicating only associative relationships. Most networks applied in psychology are undirected, such as Gaussian graphical models (GGM), Ising models, and mixed graphical models (MGM). These models are typically estimated within the framework of pairwise Markov random fields (PMRF), whereby undirected edges represent conditional dependencies between nodes; the absence of an edge implies that two nodes are conditionally independent given all other nodes in the network. Researchers select specific PMRF variants according to data type: continuous data are modeled using GGMs (Costantini et al., 2015; Lauritzen, 1996); binary data with Ising models (van Borkulo et al., 2014); and mixed data (both continuous and categorical) with MGMs (Haslbeck & Waldorp, 2015).

When defining network structure, not all nodes contribute equally. Recent research has concentrated on identifying core or "central" symptoms within psychological or psychopathological networks, with the aim of targeting these symptoms for precision interventions in mental health treatment (Berlim et al., 2021; Yang et al., 2023). To evaluate the relative importance of nodes (symptoms) within a network, researchers have introduced the notion of node centrality (Bringmann et al., 2019). Originating in social network analysis, Freeman's seminal work in 1979 delineated three centrality metrics: degree centrality, closeness centrality, and betweenness centrality. Degree centrality quantifies the number of direct connections a node has with others; closeness centrality measures how proximate a node is to all other nodes in the network; and betweenness centrality identifies nodes that serve as critical bridges within the network (Freeman & Freeman, 1979; Iacobucci et al., 2017). Nodes exhibiting the highest values on these metrics are deemed core to the network's structure. Translated into psychological networks, nodes with maximal centrality are considered optimal targets for intervention, as they represent the most influential symptoms (Ugurlu, 2022; Williams et al., 2021). Consequently, an array of centrality indices—such as strength, betweenness, closeness, and expected influence—alongside bridge centrality metrics—including bridge strength, bridge betweenness, bridge closeness, and bridge expected influence—are employed to quantify both overall and cross-community centrality of nodes (Epskamp et al., 2018; Jones et al., 2021).

1.1 The Undirected Nature of Cross-Sectional Network Analysis

Cross-sectional network analysis refers to the use of cross-sectional data to investigate psychological networks. When working with cross-sectional data, researchers typically employ undirected graphs—such as Gaussian graphical models, Ising models, and mixed graphical models—which have accounted for the vast majority of empirical network studies over the past decade (Burger et al., 2023).

However, the principal drawback of undirected graphs lies in their very definition: they do not allow researchers to infer possible causal directions between nodes within the network (Junus & Yip, 2024; Lunansky et al., 2022).

In the absence of directional information, can we rely on network centrality indices to identify core nodes? Over the past few years, many investigators have indeed done so, selecting nodes with high centrality indices as the network's core symptoms (Robinaugh et al., 2020). Yet this practice may introduce substantial errors, because centrality indices quantify structural information within the network rather than directional influences between nodes. A node with high centrality is not necessarily the optimal target for intervention.

Consider, for example, the two networks depicted in Figure 1, each of which yields identical centrality indices. In Figure 1A, node A is influenced by all other nodes but does not influence them in return, reflecting a strong in-influence effect—that is, A is easily activated by the rest of the network. By contrast, in Figure 1B, node A influences all other nodes without being influenced by them, reflecting a strong out-influence effect—i.e., A is capable of activating the rest of the network. If one were to intervene on node A in Figure 1B, the effects of that intervention might propagate to nodes B, C, D, and E; however, an intervention on node A in Figure 1A might fail to transmit its effects to the other four nodes (Wysocki et al., 2022).

This example underscores that causal relationships between symptoms are not manifest in cross-sectional network models. Therefore, identifying core symptoms or bridge symptoms within a cross-sectional network necessitates the assumption of an underlying causal process that pervades the network's structure.

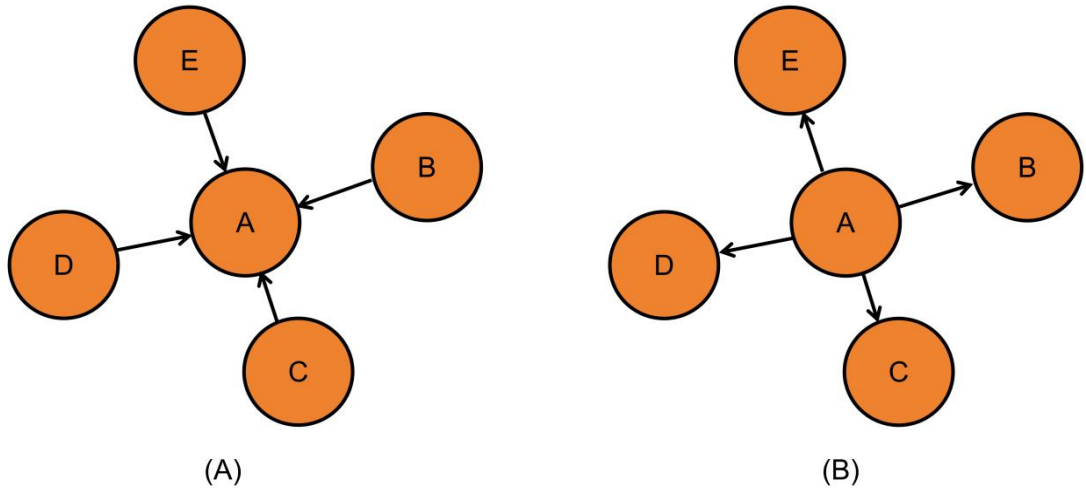

Figure 1. Schematic diagram of the symptom network with two different causal directions

1.2 NIRA: Constructing "Directed Networks" in Cross-Sectional Data

Just as cross-lagged panel models and time-series networks harness temporal information to construct directed graphs—thereby satisfying Granger's criterion that causes precede their effects (Granger, 1969)—experimental designs in psychology manipulate independent variables to observe subsequent changes in dependent variables, ensuring that the former precede the latter. This raises the question: is there an algorithm capable of inferring a "precedence" ordering among nodes in a cross-sectional network? Lunansky et al. (2022) introduced such a method, termed the nodeIdentifyR algorithm (NIRA), also known as simulated intervention or "in silico intervention."

NIRA assesses each node's causal influence on the psychological network by systematically altering its characteristics and observing downstream effects

(Lunansky et al., 2022). Concretely, one simulates an intervention by raising or lowering a node's spontaneous activation probability and then records the resulting changes across all other nodes. Repeating this procedure for each node identifies the one whose perturbation yields the greatest cumulative network effect—this node is deemed the optimal intervention target. It is important to note that, in real-world clinical contexts, symptoms rarely vanish entirely post-intervention; ethically, we aim instead to reduce symptom severity and prevalence. NIRA transcends this limitation by modulating activation probabilities in silico, thereby revealing not only the prime intervention targets (nodes whose down-regulation most attenuates network activation) but also optimal prevention targets (nodes whose up-regulation most fortifies network resilience) (Lunansky et al., 2022).

Simulated intervention in cross-sectional networks holds considerable promise. Unlike traditional reliance on centrality or bridge-centrality metrics—which may misidentify core symptoms when causal directions among symptoms are unknown—NIRA explicitly simulates causal processes to overcome this ambiguity (Lunansky et al., 2022). By imposing a hypothetical causal perturbation, NIRA can pinpoint the symptoms within a psychopathological network that are most susceptible to exacerbation or alleviation, thereby guiding real-world prevention and intervention strategies. Furthermore, it can identify bridge targets in comorbidity networks, indicating nodes whose manipulation in one disorder can ameliorate symptoms of another (Wang et al., 2025). Beyond psychopathology, this approach is broadly applicable across psychology, as it augments traditional cross-sectional analyses with simulated causality, thereby enhancing the interpretability and persuasiveness of network findings (Li et al., 2024a; Li et al., 2024b; Machado et al., 2024).

1.3 Multilevel NIRA: An Algorithm for Cross-Sectional Nested Data

Many cross-sectional datasets are inherently nested—students within schools (Raudenbush & Bryk, 1986), residents within cities (Wu et al., 2024). A defining feature of nested data is that observations are not statistically independent, violating the assumptions of single-level models such as ordinary least-squares regression or logistic regression (Cohen, 1988). Applying single-level techniques to nested data can therefore yield substantial inferential errors. Common multilevel approaches include linear mixed-effects model (LMM; Goldstein, 1986), which extend linear regression, and multilevel logistic regression model (MLRM; Wong & Mason, 1985), which extend logistic regression. These models are widely used in psychological research to address issues of nested data (O'Connell & McCoach, 2004). However, multilevel modeling is not always necessary: when between-group variance is negligible—as assessed by the intraclass correlation coefficient (ICC; DiPrete & Forristal, 1994)—ordinary least-squares regression may suffice.

Although the nodeIdentifyR algorithm (NIRA) represents a promising advance in simulated intervention techniques, it currently cannot accommodate nested data structures (Lunansky et al., 2022). Extending NIRA to a multilevel framework would not only resolve statistical modeling issues but also carry substantial practical significance, enabling identification of optimal intervention and prevention targets specific to each group. For example, if we collect anxiety measures from residents in every province and municipality of China, single-level NIRA yields only the average optimal intervention targets across all regions. A multilevel NIRA, by contrast, would allow us to select any single province or municipality and determine its region-specific optimal targets for anxiety intervention. In the following "Empirical Application" section, we implement this idea in detail and explain each step.

Building on the original simulated-intervention framework, the present study develops a multilevel NIRA algorithm. First, we introduce its two key components—multilevel logistic regression and the multilevel Ising model. We then describe the theoretical principles underlying multilevel NIRA and demonstrate its empirical application by modeling an anxiety network using data from the Psychological and Behavioral Investigation of Chinese Residents 2023 (PBICR-2023). Finally, we discuss the opportunities, challenges, and appropriate contexts for employing this novel methodology.

**2 Multilevel Binary Logistic Regression**

Prior to introducing the multilevel extension, we briefly review standard logistic regression. Logistic regression is a probability - based classification algorithm suited to data with a categorical dependent variable. Like linear regression and other supervised learning methods, it seeks to estimate a function mapping predictors X to an outcome Y, but here the predicted value represents a probability *p* rather than a continuous response. By comparing *p* to a threshold (commonly 0.5), the model produces a binary classification. Specifically, similar to linear regression, logistic regression also yields an expression of the form Y = $\beta_0 + \beta_1 X_1 + \ldots + \beta_k X_k$. However, the output of logistic regression is a probability value, so the values of $\beta_0 + \beta_1 X_1 + \ldots + \beta_k X_k$ need to be transformed into probability values $P$ within the range 0-1 using an activation function (the sigmoid function, see Equation 1). By examining this activation function expression, we find that its leftmost end approaches 0 and its rightmost end approaches 1, with its domain between (0,1). This is precisely the property required for probabilities, thus we obtain the functional expression of logistic regression, as shown in Equation 2. In Equation 2, X$\beta$ represents the dot product operation of vectors, whose value equals $\beta_0 + \beta_1 X_1 + \ldots + \beta_k X_k$.

$$\sigma(x) = \frac{1}{1 + e^{-x}} \tag{1}$$

$$P(X) = \frac{1}{1 + e^{-X\beta}} \tag{2}$$

Once the logistic model has been specified, the next task is to estimate the parameter vector *β*. The most widely used approach is maximum likelihood estimation (MLE), whose fundamental principle is to choose *β* so that the probability of observing the data—given the model—is maximized. Because we will derive and discuss the likelihood function for the multilevel binary logistic regression in a subsequent section, we do not elaborate further on MLE here. We also need to pay attention to the meaning of the slope and intercept in logistic regression, especially the intercept, which is at the core of NIRA. For example, if we want to study the impact of 17 types of adverse childhood experiences on depression, we assume that depression is a binary variable (not depressed/depressed). The slopes for the 17 types of adverse childhood experiences are $\beta_1$ to $\beta_{17}$. The meaning of the slope is that for each additional unit of an adverse childhood experience, compared to the original adverse childhood experience, the odds of developing depression increase by $exp(\beta_1), \ldots, exp(\beta_{17})$ times.

The intercept term $\beta_0$ represents the model's predicted log-odds—or, after transformation, the probability—of depression when all predictors are set to zero. As shown in Equation 4, the logistic transformation ensures that larger intercepts map to higher baseline probabilities and smaller intercepts to lower baseline probabilities. In other words, $\beta_0$ can be interpreted as the spontaneous activation probability of the outcome in the absence of any measured risk factors. This baseline–probability interpretation lies at the heart of NIRA's simulated interventions: NIRA treats $\beta_0$ as the intrinsic probability of symptom—or disorder—activation when external triggers are absent, and by artificially increasing or decreasing this probability in silico, it emulates the real-world processes of symptom exacerbation and alleviation.

However, when the data are hierarchically structured—for example, students nested within schools—applying standard logistic regression directly will severely neglect within-group effects and can lead to serious inferential errors (Nezlek, 2008). Consequently, we must adopt a multilevel logistic regression model (MLRM) to properly model such nested data (Wong & Mason, 1985). Since the present study is devoted to the development of multilevel NIRA, we provide a detailed account of the key modeling steps. The specific procedure for fitting a multilevel binary logistic regression is as follows:

Let $y_{it}$ denote the observations from the $t$-th time point ($t = 1, \ldots, T_i$) for the $i$-th group ($i = 1, \ldots, n$), and convert these observations into vector form: $y_{it} = (y_{i1}, \ldots, y_{iT_i})^\top$. Similarly, let $x_{it} = (1, x_{it1}, \ldots, x_{itp})^\top$ be the independent variable vectors corresponding to fixed effects, which include an intercept term 1 and $p$ independent variables. Let $z_{it} = (1, z_{it1}, \ldots, z_{itq})^\top$ be the independent variable vectors corresponding to random effects, which include an intercept term 1 and $q$ independent variables with random slopes. Since adding random slopes would make the model overly complex in subsequent multi-level Ising models, we only consider modeling with random intercepts here, so $z_{it} = (1)$. For binary variables (0/1 variables), given the random effect $b_i$ and independent variables $x_{it}$, assume that $y_{it}$ is conditionally independent, with mean $\mu_{it} = P(y_{it} = 1 | b_i, x_{it})$ and variance $\text{Var}(y_{it}|b_i) = \mu_{it}(1 - \mu_{it})$, expressed as

$$g(\mu_{it}) = x_{it}^T \beta + z_{it}^T b_i = \eta_{it}^{par} + \eta_{it}^{rand} \tag{5}$$

In this Equation, $g$ is a monotonic and continuously differentiable link function. For binary variables, we use the logit link, where $g(\mu_{it}) = \log \frac{\mu_{it}}{1-\mu_{it}}$; $\eta_{it}^{par} = x_{it}^T \beta$ represents the fixed effects part, including a fixed intercept and a fixed slope. The parameter vector $\beta^T = (\beta_0, \beta_1, \ldots, \beta_p)$; $\eta_{it}^{rand} = z_{it}^T b_i$ represents the random effects part. $b_i$, which follows a normal distribution, $b_i \sim N(\mathbf{0}, \mathbf{Q})$, where $\mathbf{Q}$ is a $q \times q$ covariance matrix. Since we do not consider random slopes at present, $\eta_{it}^{rand} = b_i$. Besides, An equivalent way to write it is

$$\mu_{it} = \frac{e^{\eta_{it}}}{1 + e^{\eta_{it}}} = P(y_{it} = 1 | x_{it}, b_i) \tag{6}$$

Where $\eta_{it} = \eta_{it}^{par} + \eta_{it}^{rand}$. For all groups, Equation 5 can be written as

$$g(\boldsymbol{\mu}) = \text{logit}(\boldsymbol{\mu}) = \mathbf{X}\boldsymbol{\beta} + \mathbf{b} \tag{7}$$

Where $\mathbf{X}^T = [\mathbf{X}_1^T, \dots, \mathbf{X}_n^T]$, $\mathbf{b}^T = (\mathbf{b}_1^T, \dots, \mathbf{b}_n^T)$.

After obtaining the expression for the multilevel binary logistic regression, we can likewise estimate the parameters that maximize the likelihood by constructing the corresponding likelihood function. Because the binary random variable $y_{it} \in \{0,1\}$, which follows a Bernoulli distribution, under the condition that fixed effects $\beta$ and within-group random effects $b_i$, the conditional density of the response variable $y_{it}$ can be written in the following exponential-family form:

$$f(y_{it}|\beta, b_i) = \exp\{y_{it}\eta_{it} - \log(1 + e^{\eta_{it}})\} \tag{8}$$

This component can be understood simply as the probability that the *t*-th observation value $y_{it}$ takes the value 0 or 1 in group *i*. However, knowing only the individual probabilities within a group is far from sufficient; we must next compute the group-level likelihood by multiplying together the probabilities for all observations in group *i*, yielding the following equation:

$$f(\mathbf{y}_i|\beta, b_i) = \prod_{t=1}^{T_i} \exp\{y_{it}\eta_{it} - \log(1 + e^{\eta_{it}})\} \tag{9}$$

To go further, we need to estimate the joint likelihood of all groups, which means multiplying the probabilities of each group again. At this point, we still lack an estimate for the random effect. The random effect $b_i$ follows a normal distribution, $b_i \sim N(\mathbf{0}, \mathbf{Q})$. A feasible approach is to incorporate the density $p(\mathbf{b}_i)$ of the random effect into the likelihood function. Finally, since the product form is not readily interpretable, we instead consider the log-likelihood by applying the natural logarithm to both sides of the equation, yielding the following log-likelihood function:

$$\ell(\beta, Q) = \sum_{i=1}^{n} \log \left[ \int \prod_{t} \exp\left(y_{it}\eta_{it} - \log(1 + e^{\eta_{it}})\right) p(\mathbf{b}_i) d\mathbf{b}_i \right] \quad (10)$$

Direct maximization of the likelihood function described above is prohibitively complex. Therefore, following the recommendations of Breslow and Clayton (1993), Lin and Breslow (1996), and Breslow and Lin (1995), we adopt the Penalized Quasi-Likelihood (PQL) method to reformulate the likelihood. Specifically, we augment the previous quasi-log-likelihood (i.e., the log-likelihood without the random-effects component) with a penalty term $\mathbf{b}^T Q^{-1} \mathbf{b}$ obtained via a Laplace approximation. The resulting penalized likelihood $\ell^{\mathrm{app}}(\beta, b, Q)$ is considerably easier to maximize, thereby enabling efficient estimation of the model parameters.

$$\ell^{\mathrm{app}}(\beta, b, Q) = \sum_{i,t} [y_{it}\eta_{it} - \log(1 + e^{\eta_{it}})] - \frac{1}{2}\mathbf{b}^T Q^{-1} \mathbf{b} \quad (11)$$

Similarly, understanding the intercepts and slopes in multilevel binary logistic regression is critically important—especially the random-intercept component, which underpins the statistical framework of our multilevel NIRA. Returning to the example introduced above, we aim to examine the impact of 17 adverse childhood experiences on depression, using data collected from *i* schools, each with *t* students. Note that students' depression scores are nested within schools; as before, applying a standard logistic regression would not adequately disentangle between-school (Level-2) effects from within-school (Level-1) effects. In this context, we again treat depression as a binary outcome (non-depressed vs. depressed) and specify fixed slopes $\beta_1$ through $\beta_{17}$ for the 17 adverse experiences. These fixed slopes mean that, across all schools, a one-unit increase in any given adverse experience is associated with an average $exp(\beta_1), \ldots, exp(\beta_{17})$-fold increase in the odds of depression, see Equation 12. By contrast, random slopes would allow each school to exhibit its own idiosyncratic odds ratio for each adverse experience; however, this study does not include random slopes, so we do not discuss them further.

$$\theta = \frac{exp(\beta_0 + \beta_1(X_{it1} + 1))}{exp(\beta_0 + \beta_1(X_{it1}))} = exp(\beta_1) \qquad (12)$$

Our focus is on the intercept component of multilevel binary logistic regression. As noted above, the intercept in a standard logistic regression represents the probability of the outcome when all predictors are set to zero. In nested data structures, this intercept is decomposed into a fixed intercept $\beta_{01}$ and a random intercept $\beta_{0i}$, see Equation 13. The fixed intercept denotes the average probability of depression (i.e., the mean baseline activation probability) across all schools when all 17 adverse childhood experiences are zero (i.e., no adverse experiences). The random intercept, by contrast, allows this baseline activation probability to vary by school—for example, School A might have a 10 % baseline probability of depression, whereas School B has only 4 %. The core innovation of multilevel NIRA interventions lies in this decomposition: by isolating the random intercept, we can simulate either intensification or mitigation interventions for each school by shifting its random intercept up or down by two standard deviations and then observing the resulting changes in the network. Likewise, by applying the same ±2-SD shift to the fixed intercept, we can examine the average simulated intervention effect across all schools.

$$P(X) = \frac{e^{\beta_0}}{1 + e^{\beta_0}} = \frac{e^{\beta_{01} + \beta_{0i}}}{1 + e^{\beta_{01} + \beta_{0i}}} \qquad (13)$$

## 3 Multilevel Ising model

Building upon the multilevel binary logistic regression introduced earlier, we can further extend the model to a multilevel Ising framework. To facilitate readers' understanding, we begin with a conceptual introduction to the standard Ising model and then gradually transition to the multilevel Ising model using the Ising-based formulation. To avoid terminological confusion, we clarify here that the terms "node" and "variable" are used interchangeably throughout the following text; in the context of network analysis, a "variable" is also referred to as a "node".

The Ising model originated in physics and was initially proposed to explain phase transitions in ferromagnetic materials—specifically, the phenomenon where a magnet loses its magnetism when heated above a certain critical temperature, and regains magnetism upon cooling below that temperature (Ising, 1925). The model describes the interaction between the states of particles connected within a network. As illustrated in Figure 2, this network comprises n nodes, with edges representing interactions between them. Each node corresponds to a particle's spin state $x_i$, where $i$ = 1, ..., n, and each spin is restricted to pointing either "up" (↑) or "down" (↓). In this model, "↑" is encoded as $x_i = +1\$$ and "↓" as $x_i = -1\$$. More specifically, Equation 14 is used to describe a nearest-neighbor network of binary random variables, namely the Ising network. Notably, Equation 14 is a transformation of Equation 2. Typically, in psychological applications, we are interested in the probability that a given variable takes the value 1. When we set the variable value to 1, Equation 14 becomes equivalent to Equation 2, underscoring that the Ising network is essentially a form of logistic regression.

$$P_\theta(x_j|x_{\setminus j}) = \frac{\exp[\tau_j x_j + x_j \sum_{k \in V_{\setminus j}} \beta_{jk} x_k]}{1 + \exp[\tau_j + \sum_{k \in V_{\setminus j}} \beta_{jk} x_k]} \quad (14)$$

Where $\tau_j$ and $\beta_{jk}$ are the threshold of nodes and the interaction parameter of paired nodes, respectively. V \j represents other variables except variable j. Correspondingly, the expression of multilevel Ising model can be written as

$$P(x_{itj}|x_{it,-j}, b_i) = \frac{\exp\left[(\tau_j + b_{ij})x_{itj} + x_{itj} \sum_{k \in V_{\setminus j}} \beta_{jk} x_{itk}\right]}{1 + \exp\left[(\tau_j + b_{ij}) + \sum_{k \in V_{\setminus j}} \beta_{jk} x_{itk}\right]} \quad (15)$$

The biggest difference from equation 6 is the inclusion of a random intercept $b_{ij}$, which means that for each variable *j,* each group *i* has its own specific intercept $b_{ij}$.

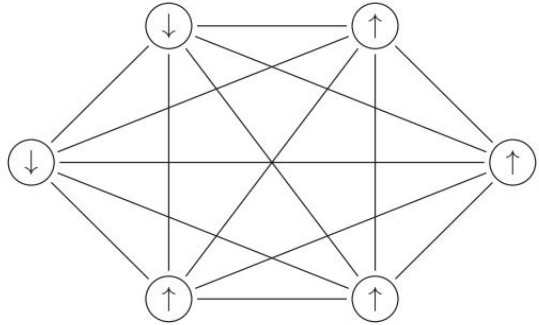

Figure 2. Schematic diagram of Ising network

The Ising model has garnered considerable attention for its structural simplicity combined with its capacity to capture complex phenomena that emerge from the interactions among nodes (Stutz & Williams, 1999). Today, this model is widely applied across numerous disciplines, including biology (Fierst & Phillips, 2015), sociology (Galam, Gefen, & Shapir, 1982; Galam & Moscovici, 1991), and psychology (Marsman et al., 2018; van Borkulo, Epskamp, & Robitzsch, 2014). In psychopathology research, the Ising model is particularly well-suited due to its alignment with network theory, and it is frequently employed to model symptom networks (Lunansky et al., 2022). In such networks, symptoms are typically coded as binary variables—0 indicating the absence of a symptom and 1 indicating its presence—thereby making the binary structure of the Ising model a natural fit. Building upon this framework, van Borkulo and colleagues (2014) developed the eLasso method to further improve the estimation of Ising networks, as shown in Figure 3.

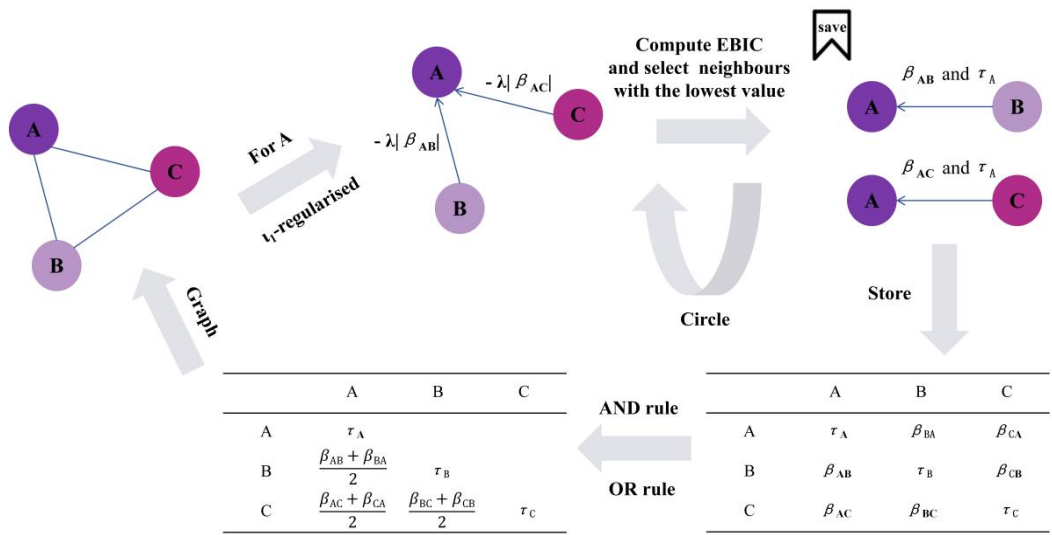

Figure 3. Construction process of Ising network

Although the construction of the single-level Ising model is not the primary focus of this study, the procedure for building multilevel Ising model closely mirrors that of the single-level model. Thus, we will elaborate on the construction process of multilevel Ising model based on the aforementioned framework. For illustrative purposes, consider a multilevel Ising network composed of three nodes (i.e., variables), as shown in Figure 3. The first step involves fitting a multilevel binary logistic regression model for each variable, where the outcome variable $X_i$ is the node of interest, and the predictor variables are all other remaining nodes. In our example, this means constructing three separate multilevel binary logistic regression models: (1) one with A as the outcome and B and C as predictors, (2) one with B as the outcome and A and C as predictors, and (3) one with C as the outcome and A and B as predictors. Figure 3 illustrates only the model where A is the outcome variable. For the sake of clarity, the modeling steps discussed below will refer to a single instance of multilevel binary logistic regression. However, it is important to note that in actual practice, this procedure must be repeated n times, where n is the total number of variables included in the multilevel Ising model.

It is important to note that in order to control for spurious edges that may arise from repeated regressions and to identify the most relevant predictors, L1 regularization is required (Meinshausen & Bühlmann, 2006; Ravikumar, Wainwright, & Lafferty, 2010; Groll & Tutz, 2014). Specifically, L1 regularization involves adding a penalty term $\lambda \sum_{i=1}^{p} |\beta_i|$ to the original likelihood function based on Equation 11. Following the approach proposed by Breslow and Clayton (1993), the PQL (Penalized Quasi-Likelihood) function incorporating the L1 regularization penalty is formulated as shown in Equation 16. Here, λ is the regularization parameter that governs the strength of the L1 penalty. Different values of λ yield models of varying complexity. In other words, for each variable, *n* models are generated corresponding to *n* values of λ. Figure 3 provides a more intuitive illustration of how L1 regularization is incorporated into the modeling process.

$$l^{\text{pen}}(\boldsymbol{\beta}, \boldsymbol{b}, \boldsymbol{Q}) = l^{\text{pen}}(\boldsymbol{\delta}, \boldsymbol{Q}) = l^{\text{app}}(\boldsymbol{\beta}, \boldsymbol{b}, \boldsymbol{Q}) - \lambda \sum_{i=1}^{p} |\beta_i| \qquad (16)$$

Where we define $\boldsymbol{\delta}^T = (\boldsymbol{\beta}^T, \boldsymbol{b}^T)$.

Given $\boldsymbol{Q}$, the above optimization problem can also be simplified as

$$\hat{\boldsymbol{\delta}} = \underset{\delta}{\operatorname{argmax}}\, l^{\text{pen}}(\boldsymbol{\delta}, \boldsymbol{Q}) = \underset{\delta}{\operatorname{argmax}} \left[ l^{\text{app}}(\boldsymbol{\delta}, \boldsymbol{Q}) - \lambda \sum_{i=1}^{p} |\beta_i| \right] \qquad (17)$$

After constructing the complete likelihood function, the next step is to maximize it. To achieve this, we adopted the approach proposed by Groll and Tutz (2014), specifically utilizing the gradient ascent algorithm to maximize the penalized likelihood function presented in Equation 16. To aid readers in better understanding this process, it is necessary to introduce the fundamentals of the gradient ascent algorithm. As vividly illustrated in Figure 4, the essence of gradient ascent is analogous to mountain climbing: in order to reach the summit, one must not only determine the correct direction to ascend but also decide the appropriate step size. Returning to our penalized likelihood function $l^{\text{pen}}(\boldsymbol{\beta}, \boldsymbol{b}, \boldsymbol{Q})$, our objective is to maximize this function and obtain the parameter estimates for $\beta$. While additional parameters must also be estimated in practice, we use $\beta$ here as a representative example to facilitate comprehension. Imagine that we are currently at point A on the slope. At this point, we do not know whether we should move upward or downward. If we should move upward, in which direction exactly should we proceed? Should we ascend at a 45-degree angle? Or perhaps at 60 degrees?

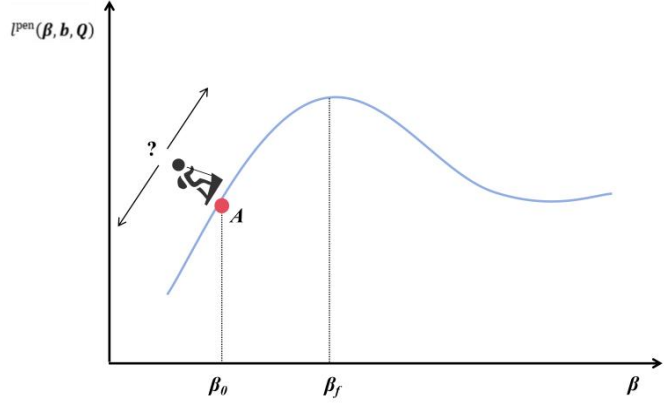

Figure 4. Schematic diagram of gradient ascent algorithm

To determine the appropriate direction of movement, we first need to compute the partial derivative of the penalized likelihood function with respect to the parameter $\beta$, that is, $\frac{\partial l^{\text{pen}}(\boldsymbol{\beta},\boldsymbol{b},\boldsymbol{Q})}{\partial \boldsymbol{\beta}}$. The collection of partial derivatives with respect to all parameters constitutes the gradient, which is a vector indicating the direction of the steepest ascent. Specifically, the gradient points in the direction of the greatest directional derivative. In the context of our example at point A, the directional derivative can be interpreted as the rate of increase of the function in a specific direction, and in our likelihood function, it is formally expressed as:

$$l'_{\text{pen}}(\boldsymbol{\delta}; \mathbf{v}, \boldsymbol{Q}) = \lim_{t \downarrow 0} \frac{1}{t}\left(l^{\text{pen}}(\boldsymbol{\delta} + t\mathbf{v}, \boldsymbol{Q}) - l^{\text{pen}}(\boldsymbol{\delta}, \boldsymbol{Q})\right) \tag{18}$$

Where $\mathbf{v}$ is used to indicate our direction, $\mathbf{v} \in \mathbb{R}^{p+nq}$.

Having determined the direction, the next critical question is: how far should we move along this direction? This is essential because if the step size is too large, we risk overshooting the optimal point $\beta_f$, thereby missing the maximum of the function. Therefore, it is necessary to define an upper bound for the step size to ensure stable convergence. Conceptually, the product of the direction and the step size yields the update magnitude. By adding this update to the current parameter estimates, we obtain new estimates. With these new values, we repeat the procedure iteratively until convergence at the maximum of the function is achieved and the parameters are fully updated. Following the approach proposed by Groll and Tutz (2014), we arrive at the parameter update formulation of the penalized likelihood function, see Equation 19. In this formulation, $t$ represents the step size mentioned above, and $\mathbf{s}^{\text{pen}}(\hat{\boldsymbol{\delta}}, \boldsymbol{Q})$ denotes the gradient vector. For the detailed mathematical expressions of both the gradient and the step size, readers are referred to the original work by Groll and Tutz (2014).

$$\begin{aligned} l^{\text{pen}}(\hat{\boldsymbol{\delta}} + t\mathbf{s}^{\text{pen}}(\hat{\boldsymbol{\delta}}, \boldsymbol{Q}), \boldsymbol{Q}) \quad &\approx l^{\text{pen}}(\hat{\boldsymbol{\delta}}, \boldsymbol{Q}) \\ &\quad + t l'_{\text{pen}}(\hat{\boldsymbol{\delta}}; \mathbf{s}^{\text{pen}}(\hat{\boldsymbol{\delta}}, \boldsymbol{Q}), \boldsymbol{Q}) \\ &\quad + 0.5 t^2 l''_{\text{pen}}\left(\hat{\boldsymbol{\delta}}; \mathbf{s}^{\text{pen}}(\hat{\boldsymbol{\delta}}, \boldsymbol{Q})\right) \end{aligned} \quad (19)$$

For each penalization parameter λ, a corresponding likelihood function can be computed and maximized using the gradient ascent algorithm to obtain the associated parameter estimates. Upon completing this first step, a critical question arises: Which value of λ yields the optimal model? This question was aptly addressed by Chen and Chen (2008), who proposed the use of the Extended Bayesian Information Criterion (EBIC) to select the best model. EBIC is an adjusted information criterion particularly suited for settings involving a large number of candidate features (i.e., potential predictors). It serves as a stopping rule in model selection: when the EBIC value reaches its minimum, it indicates that the optimal subset of features has been identified. Therefore, among the n models generated for the n different values of λ, the one with the smallest EBIC is regarded as the optimal model. In the present study, the EBIC is defined as follows:

$$\text{EBIC}_\gamma = -2l^{\text{pen}}(\boldsymbol{\delta}, \boldsymbol{Q}) + |M|\ln(n) + 2\gamma \ln \binom{p}{|M|} \quad (20)$$

Where *M* is the number of fixed-effect predictors, n is the number of observations of the binary outcome variable, $\gamma \in \{0,1\}$ is a tuning parameter that controls the strength of the additional complexity penalty, and *p* is the total number of candidate fixed-effect predictors. Since each binary variable is used in the multilevel Ising model, we define $p = M$.

After identifying the optimal model, we proceed to the third step: extracting the regression coefficients (fixed slopes), fixed intercepts, and random intercepts from this model. Taking the multilevel binary logistic regression model with variable A as the outcome (as shown in Figure 3) as an example, note that Figure 3 represents a single-level Ising model, thus only the fixed slopes $\beta_{AC}$, $\beta_{AB}$ and the fixed intercept $\tau_A$ are extracted.

However, in the multilevel Ising model, particular attention is given to the random intercepts $b_{iA}$, as they reflect group-specific spontaneous activation probabilities. Therefore, it is necessary to extract the random intercepts as well. As illustrated in Figure 5, the extraction of parameters is divided into two parts. The first part pertains to the fixed effects: fixed slopes (regression coefficients) and the fixed intercept are entered into a parameter matrix, with the fixed intercept placed on the diagonal and the regression coefficients placed in the off-diagonal elements. The second part concerns the random effects: similarly, fixed slopes and random intercepts are placed into a parameter matrix, but in this case, the random intercepts are positioned along the diagonal and the regression coefficients off-diagonally. Through this process, both fixed-effect and random-effect parameter matrices are constructed.

**Fixed Intercepts**

|   | A | B | C |
|---|---|---|---|
| A | $\tau_A$ | $\beta_{BA}$ | $\beta_{CA}$ |
| B | $\beta_{AB}$ | $\tau_B$ | $\beta_{CB}$ |
| C | $\beta_{AC}$ | $\beta_{BC}$ | $\tau_C$ |

AND rule → OR rule

|   | A | B | C |
|---|---|---|---|
| A | $\tau_A$ | | |
| B | $\frac{\beta_{AB}+\beta_{BA}}{2}$ | $\tau_B$ | |
| C | $\frac{\beta_{AC}+\beta_{CA}}{2}$ | $\frac{\beta_{BC}+\beta_{CB}}{2}$ | $\tau_C$ |

**Random Intercepts**

|   | A | B | C |
|---|---|---|---|
| A | $b_{iA}$ | $\beta_{BA}$ | $\beta_{CA}$ |
| B | $\beta_{AB}$ | $b_{iB}$ | $\beta_{CB}$ |
| C | $\beta_{AC}$ | $\beta_{BC}$ | $b_{iC}$ |

AND rule → OR rule

|   | A | B | C |
|---|---|---|---|
| A | $b_{iA}$ | | |
| B | $\frac{\beta_{AB}+\beta_{BA}}{2}$ | $b_{iB}$ | |
| C | $\frac{\beta_{AC}+\beta_{CA}}{2}$ | $\frac{\beta_{BC}+\beta_{CB}}{2}$ | $b_{iC}$ |

Figure 5. Intercept extraction in the multilevel Ising model

It is important to note that upon completing Steps 2 and 3, we have only obtained a multilevel binary logistic regression model with a single variable (e.g., variable A) as the outcome in the multilevel Ising framework. Therefore, Steps 2 and 3 must be repeated for all remaining variables (in our example, variables B and C).

At this point, for any pair of nodes $j$ and $k$, we obtain both $\beta_{jk}$ (i.e., the regression coefficient of node $j$ predicting node $k$) and $\beta_{kj}$ (i.e., the coefficient of node $k$ predicting node $j$). This means that between any two nodes in the network, there exist two directed regression coefficients. The process of determining whether an edge exists between these two nodes, and how to assign its weight based on the regression coefficients, constitutes Step 5. In Step 5, we determine whether an edge exists between nodes $j$ and $k$ by applying one of two commonly used rules: the AND rule or the OR rule (Meinshausen & Bühlmann, 2006; Ravikumar, Wainwright, & Lafferty, 2010). Specifically, under the AND rule, an edge is established between nodes $j$ and $k$ only if both $\beta_{jk}$ and $\beta_{kj}$ are non-zero; the edge weight is then defined as the average of these two coefficients. In contrast, under the OR rule, the presence of a non-zero value in either $\beta_{jk}$ or $\beta_{kj}$ is sufficient to define an edge between the two nodes; the edge weight is taken to be the non-zero coefficient among the two. At this stage, both the fixed-effect and random-effect parameter matrices have been fully updated.

The AND and OR rules are primarily applied to the regression coefficients in order to determine the presence and weight of edges in the network. In the following section, we shift focus to the intercepts in the multilevel Ising model. In line with multilevel binary logistic regression, and following the perspective of van Borkulo et al. (2014), intercepts are also referred to as threshold parameters, representing a node's intrinsic tendency to activate (i.e., take the value 1) in the absence of influence from adjacent variables. In other words, the intercept quantifies the probability that a given variable transitions from 0 to 1 when all other variables are held at 0 (i.e., absent). In the multilevel modeling context, intercepts are divided into fixed intercepts and random intercepts. According to the above definition, the fixed intercept can be interpreted as the average probability that a variable changes from 0 to 1 across all groups, given that all other variables are set to 0. Conversely, the random intercept reflects the group-specific probability that a variable switches from 0 to 1 under the same condition, thereby capturing spontaneous activation tendencies that vary across groups.

To illustrate with a simple example, as shown in Figure 3, suppose we collect data on cold symptoms among students from different schools. Variables A, B, and C represent symptoms of the common cold—nasal congestion, sore throat, and coughing, respectively (Andrewes, 1949). In this case, the network represents a multilevel network of cold symptoms. Taking variable A as an example, the fixed intercept $\tau_A$ indicates the average probability across all schools that a student exhibits nasal congestion (i.e., a transition from 0 = absent to 1 = present) when neither sore throat nor coughing is present. The random intercept $b_{iA}$ reflects the school-specific probability that a student in a given school (e.g., School A) experiences nasal congestion under the same conditions.

As previously discussed, the core mechanism of Multilevel NIRA lies in modifying the intercept terms of multilevel binary logistic regression models to simulate real-world scenarios of symptom exacerbation or alleviation. Given that the multilevel Ising network is essentially composed of a set of multilevel binary logistic regression models, Multilevel NIRA can naturally be applied to conduct interventions on each variable within the network. Specifically, Multilevel NIRA operates by increasing or decreasing the intercepts (i.e., fixed and/or random intercepts) of the logistic regression models for each of the n variables, thereby simulating an increase or decrease in the spontaneous activation probability (i.e., the average or group-specific baseline probability) of each variable. This procedure generates n+1 networks (with the "+1" referring to the original, unaltered network). By comparing the total scores across these networks, researchers can identify which variable serves as the optimal target for prevention or intervention. Due to the structure of multilevel Ising networks, they are particularly well suited for integration with the Multilevel NIRA framework. This approach allows researchers to examine which variables play a relatively central or influential role within the overall network. In the context of psychopathology, this is especially valuable for identifying critical future targets for prevention or intervention, and it may also offer preliminary insights into underlying causal relationships. Detailed procedures for increasing or decreasing intercept terms are outlined in the following section.

**4 Multilevel NIRA**

According to current research practices, the implementation of NIRA generally involves four key procedures: "constructing the Ising network," "simulating the generation of the original sample," "simulating interventions and generating post-intervention samples," and "intergroup comparison of total mean scores" (Wang et al., under review). The implementation of multilevel NIRA largely parallels that of the standard NIRA procedure. In the following sections, we elaborate on each component.

4.1 Constructing the Multilevel Ising Network

This process was described in detail in the previous section; here we provide a brief recap. The first step is to ensure that the data entered into the multilevel Ising model are binary (0/1), and that appropriate centering operations are conducted according to the specific research objectives. The second step involves performing $N$

L1-regularized multilevel logistic regressions ($N$ being the number of variables). In the third step, for each variable, the Extended Bayesian Information Criterion (EBIC) values are computed for $k$ models (where $k$ is the number of penalty parameters $\lambda$), and the optimal model is selected (Chen & Chen, 2008; Foygel & Drton, 2011). In the fourth step, the regression coefficients and intercepts from the optimal model are extracted. After applying either the AND-rule or the OR-rule to the regression coefficients, a weight matrix is constructed. Meanwhile, the intercepts are separated into fixed and random threshold parameters.

Specifically, the weight matrix refers to the M × M edge weight matrix of the multilevel Ising network, where each cell represents the regression coefficient from the multilevel logistic regression for a pair of variables, and $M$ is the number of nodes (i.e., variables). The fixed threshold parameter vector is an M × 1 vector in which each cell contains the fixed threshold (intercept) for each node, derived from the multilevel logistic regression. The random threshold parameter matrix is an M × $i$ matrix, where each cell corresponds to the intercept for a given node in a specific group, and $i$ is the number of groups. For example, in the case of adverse childhood experiences across schools, this would result in a 17 × 3 matrix—assuming 17 variables and 3 schools—indicating the node-specific intercepts for each school.

4.2 Simulating the generation of the original sample

The second step of multilevel NIRA involves simulating a new batch of original samples based on the original weight matrix and threshold parameters. The typical sample size for such simulations is set at 5,000 (Lunansky et al., 2022; F. Wang et al., 2025). For example, based on the original dataset, we may have already obtained the weight matrix and threshold parameters from a multilevel Ising model corresponding to adverse childhood experiences. Using these original parameters, we can simulate responses from 5,000 participants on the adverse childhood experiences scale, with all responses being binary (0/1) variables. At this point, two important questions arise: first, how is the simulation conducted? Second, given that the original sample already exists, why is it necessary to simulate a new original sample based on the same weight matrix and threshold parameters?

Let us begin with the first question—how to simulate data based on the weight matrix and threshold parameters of the multilevel Ising model. The method used here is Metropolis–Hastings sampling. Before delving into the specifics of Metropolis–Hastings, it is essential to understand that for Ising model, in addition to the probability function expression of the Ising model we presented earlier, there exists an alternative formulation based on its energy function. In fact, this energy-based expression originates from the Ising model's original theoretical definition. These two expressions are mathematically interchangeable. According to Ising's original definition (Ising, 1925), the energy function of the Ising model can be written as:

$$H(s) = -\frac{1}{2}\sum_{i=1}^{N}\sum_{j=1}^{N} w_{ij}\, s_i s_j - \sum_{i=1}^{N} \theta_i\, s_i \tag{21}$$

Where $w$ represents the weight matrix, $s = (s_1, \ldots, s_N)$ is the state vector of the system, with each variable $s_i \in \{0,1\}$, N is the number of variables, and $\theta_i$ is the threshold parameter (intercept) for node *i*. Note that this expression corresponds to the single-level Ising model, and therefore the thresholds here refer to fixed intercepts. The corresponding probability distribution of the Ising model is given by:

$$P(X = s) = \pi(s) = \frac{e^{-\beta H(s)}}{Z} \tag{22}$$

Where $\beta$ is the inverse temperature parameter, typically set to 1 in psychological research, as is also the case in the present study. Interestingly, this opens up avenues to further explore the concept of network temperature (Grimes et al., 2025), which we will discuss in a subsequent section. The term $Z$ denotes the partition function.

When extending to the multilevel Ising model, if we wish to examine random effects, the random intercepts are incorporated into the model as random threshold parameters. Accordingly, the energy function for the Ising model with random effects is expressed as:

$$H(s) = -\frac{1}{2}\sum_{i=1}^{N}\sum_{j=1}^{N} w_{ij} s_i s_j - \sum_{i=1}^{N} b_{(g)i} s_i \tag{23}$$

Here, *g* denotes the *g*-th group. Since the threshold parameter supports only an M×1 vector as input, it implies that if we aim to construct an Ising model for each group, we should build *g* separate Ising models rather than directly inputting the entire random intercept matrix into the model. For instance, we first extract the intercepts corresponding to School A to construct Model A, then extract the intercepts for School B to construct Model B, and so on until all groups have been traversed. The formulation for fixed intercepts is similar to that for random intercepts and will not be elaborated here.

Once we have established the energy function, we proceed to simulate samples using the Metropolis–Hastings (MH) sampling method. A single iteration of the MH sampling procedure involves the following steps:

Step 1: Proposal. Randomly select one node $k \in \{1, \ldots, N\}$ ($N$ is the number of nodes) and flip its current binary state. Let the proposed state vector be denoted as:

$$s_i' = \begin{cases} 1 - s_k, i = k, \\ s_i, i \neq k \end{cases} \quad (24)$$

For example, assume we have 4 nodes such that the current state vector is $s = (s_1, s_2, s_3, s_4) = (0,1,0,1)$. If we randomly select node 3, we flip its value: 1−0=1. The updated (proposed) state becomes $s = (s_1, s_2, s_3', s_4) = (0,1,1,1)$, with all other values unchanged.

But does this sampled proposal conform to our target distribution $\pi(s)$? This is where Steps 2 through 4 come into play for validation.

Step 2: Compute the energy change $\Delta H$ resulting from the flip, as shown in Equation (25).

$$\begin{aligned} \Delta H &= H(s') - H(s) \\ &= (2s_k - 1)\left(\sum_{j=1}^{N} w_{kj} s_j + b_{(g)k}\right) \end{aligned} \quad (25)$$

Step 3: Determine the acceptance probability $\alpha$, which is calculated based on the ratio of the target distribution probabilities, as shown in Equation (26).

$$\alpha = \min\{1, \frac{\pi(s')}{\pi(s)}\} \quad (26)$$

Where $\frac{\pi(s')}{\pi(s)}$ can be equivalent to

$$\frac{\pi(s')}{\pi(s)} = \frac{\exp(-\beta H(s'))}{\exp(-\beta H(s))} = \exp(-\beta[H(s') - H(s)]) \qquad (27)$$
$$= \exp(-\beta\Delta H)$$

So

$$\alpha = \min\{1, \frac{\pi(s')}{\pi(s)}\} = \min\{1, \exp(-\beta\Delta H)\} \qquad (28)$$

Step 4: Decide whether to accept the proposed flip. First, generate a uniform random number $u \sim U(0,1)$. Then, determine acceptance by comparing $u$ and $\alpha$: if $u \leq \alpha$, accept the proposed state; otherwise, retain the current state.

$$s \leftarrow \begin{cases} s', & \text{if } u \leq \alpha, \\ s, & \text{otherwise.} \end{cases} \qquad (29)$$

At this point, we have fully described a single iteration of the Metropolis–Hastings (MH) sampling procedure (hereafter referred to as a single-step MH). In our study, the sample simulation proceeds as follows:

Step 1: We begin by initializing a starting binary variable vector $s^{(0)}$, where each element takes a value of either 0 or 1. For example, for the vector representing Adverse Childhood Experiences (ACEs), we might arbitrarily initialize it as $s_{ACE}^{(0)} = (s_1, \ldots, s_{17}) = (0, \ldots, 1)$.

Step 2: We then enter the burn-in phase, which aims to bring the initially selected random variable vector $s^{(0)}$ into alignment with the target distribution. Specifically, we perform *B=10\*N* single-step MH updates (where *N* is the number of variables). The resulting variable vector is denoted as $s^{(B)}$. It is important to note that this stage serves only to establish a suitable starting point for subsequent sample generation. Thus, intermediate values are not recorded, and only the final state $s^{(B)}$ is retained.

Step 3: Using $s^{(B)}$ as the initial state, we perform $N$ single-step MH updates to generate the first simulated sample $s^{(1)}$. This is conceptually equivalent to simulating a participant's responses to all 17 items on the ACE questionnaire, resulting in a binary vector such as $s^{(1)}_{ACE} = (s_1, \dots, s_{17}) = (1, \dots, 1)$. Subsequently, using $s^{(1)}$ as the new starting point, we again perform NN single-step MH updates to generate the second simulated sample $s^{(2)}$, and so on. This process is repeated iteratively until a total of 5,000 samples are generated.

Next, we address the second question: Given that we already possess the original sample, why do we need to simulate a new "original" sample based on the same weight matrix and threshold parameter vector?

The answer to this question is closely related to Section 4.3. In essence, the simulated intervention described in Section 4.3 involves generating a post-intervention sample by modifying the threshold parameter vector—either alleviating or aggravating symptoms—while keeping the weight matrix constant. To evaluate the effectiveness of such a simulated intervention, researchers must compare the post-intervention sample with a non-intervened counterpart (as discussed in Section 4.4).

However, using the original empirical sample as the control group in this comparison introduces two major methodological concerns. First, since the original sample did not undergo the same simulation process as the post-intervention sample, any observed differences between the two groups may be confounded—attributable either to the simulation procedure itself or to the intervention manipulation, making causal interpretation ambiguous. Second, the sample size of the original dataset may differ from that of the simulated post-intervention sample, which can further bias group comparisons or affect statistical power. Therefore, to avoid these issues, we simulate a new "original" sample based on the same weight matrix and the same threshold parameter vector as the post-intervention simulation. This ensures consistency in the simulation process and equivalence in sample size, providing a valid baseline for assessing the intervention's effect.

4.3 Simulating interventions and generating post-intervention samples

The third step in multilevel NIRA involves simulating targeted alleviating (or aggravating) interventions by systematically decreasing (or increasing) the threshold parameters of each node in the multilevel Ising network. Based on these adjusted threshold vectors, and in combination with the original weight matrix, new post-intervention samples are generated to reflect the hypothetical effects of such interventions, as illustrated in Figure 6. To emphasize the advantages of the multilevel model, we focus here on the procedures involving random threshold parameters (i.e., random intercepts); the steps involving fixed threshold parameters are nearly identical and will not be repeated.

This step constitutes the core innovation of both NIRA and multilevel NIRA, as it encapsulates a novel conceptualization of psychological intervention. Specifically, in the NIRA framework, an intervention targeting a particular variable is operationalized as a change in its spontaneous activation probability. That is, alleviating (or aggravating) a symptom within a psychological network reduces (or increases) the likelihood that the symptom activates in the absence of influence from other variables. Clearly, this aligns well with the theoretical definition of psychological intervention. In the context of an Ising network, a symptom's spontaneous activation probability corresponds to the intercept term in a logistic regression model—referred to here as the threshold parameter. In a multilevel Ising network, this intercept can be decomposed into a fixed component and a random component, with the latter capturing group-specific deviations in spontaneous activation. Therefore, a simulated alleviating or aggravating intervention targeting a specific node can be implemented by decreasing or increasing only that node's threshold parameter, while keeping the threshold parameters for all other nodes unchanged.

Figure 6 demonstrates this process within a multilevel Ising framework. For a given group *i*, we can generate a post-intervention sample of 5,000 individuals by altering the random threshold parameter (represented by the 0/1 bar in the figure) for node "A"—either increasing or decreasing its value—while keeping the random threshold parameters for the other two nodes constant. Using the previously obtained weight matrix and the newly modified random threshold vector, MH sampling is applied to simulate this group-level intervention outcome. Each resulting data point represents a simulated individual in group *i*, with binary responses for all three variables (nodes). This process is repeated for each remaining node in the multilevel Ising network, ultimately yielding three distinct post-intervention samples for group *i*, each reflecting a targeted intervention on a different node. It is important to emphasize that this simulation is always conducted within group *i*, as we can only input one group-specific threshold vector into the model at a time. As previously noted, the full random threshold matrix cannot be used in a single simulation.

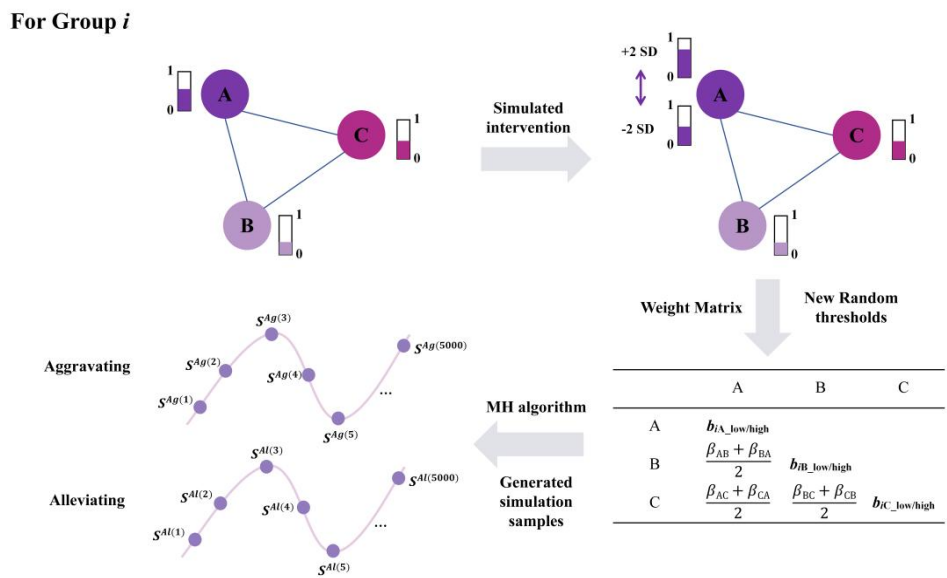

Figure 6. Multilevel NIRA simulation sample generation process

Another important consideration is the magnitude by which each node's threshold parameter should be increased or decreased during the simulated intervention. In line with the original NIRA framework, the intervention strength in multilevel NIRA is operationalized by adjusting threshold parameters in terms of their standard deviations. Specifically, the most common approach is to apply a shift equivalent to two standard deviations from the original threshold value to represent the intensity of the intervention (Lunansky et al., 2022; F. Wang et al., 2025).

Prior studies have shown that when the intervention strength is reduced from two standard deviations to one, the predictive efficacy of NIRA also diminishes accordingly (Lunansky et al., 2022). This is because a weaker intervention naturally exerts a smaller influence on the network structure. The use of standard deviation as a scaling metric provides a crucial advantage: the intervention strength is anchored in the empirical distribution of the data. In other words, the adjustment magnitude is informed by the variability in threshold parameters across all nodes in the network, rather than being an arbitrary fixed value. This approach allows for greater adaptability across different datasets and symptom networks. Conversely, using a fixed value (e.g., +1) to represent intervention strength may result in inconsistent intervention effects across nodes. This is because the same numerical shift can have markedly different impacts depending on whether a node's original threshold is relatively high or low. By defining intervention strength in terms of standard deviation, the model ensures a more consistent and statistically grounded intervention effect that reflects the underlying distribution of threshold parameters in the network.

4.4 Intergroup Comparison of Total Mean Scores

The fourth step of the multilevel NIRA procedure involves identifying the most influential node(s) within the network by comparing the simulated original sample with samples generated after simulated interventions on each individual node. Specifically, for each simulated sample—including the simulated original sample—the total score of each simulated participant is calculated by summing their responses across all variables (nodes). Then, the mean total scores of participants under each condition are visualized, and independent samples *t*-tests are conducted to evaluate the statistical significance of differences between the mean total score of the simulated original sample and those of the intervention-based samples. To control for Type I error, *p*-values are corrected for multiple comparisons (Wang et al., 2025).

In this section, it is necessary to distinguish between two types of network structures: single-subnetwork (hereafter, "single-network") and double-subnetwork (hereafter, "double-network") models, as illustrated in Figures 7 and 8. Figure 7 depicts a single-network representing Adverse Childhood Experiences (ACEs), comprising 17 ACE items. In contrast, Figure 8 shows a double-network consisting of anxiety (7 items) and depression (9 items). Although all previous steps in the multilevel NIRA procedure were demonstrated using the single-network as an example, both network structures follow the same procedure up to this point. The distinction emerges only in this final step, due to differences in research aims.

For example, in the case of a single-network, the research goal is to identify which of the 17 ACE events is the most central or impactful. Therefore, the mean total score of the simulated original sample is compared with the mean total scores from 17 intervention-based samples—each representing a simulated intervention on one specific node. As illustrated in Figure 9A, for a given group $i$, if a mitigation intervention is applied to the CUR node (by reducing its random intercept by two standard deviations), the resulting sample exhibits the lowest mean total score. This suggests that the CUR node is the optimal target for intervention. Similarly, aggravation interventions can be used to identify the best prevention targets in the network.

However, for a double-network, the research aim shifts. Rather than identifying the most central depressive symptom, the goal is to determine which depression node most strongly influences anxiety. In this case, interventions are applied to each depression node individually, but the outcome of interest is the total score of the anxiety subnetwork. As illustrated in Figure 9B, the UW node within the depression subnetwork emerges as the optimal target for intervention in reducing anxiety, indicating its critical bridging role between the two constructs.

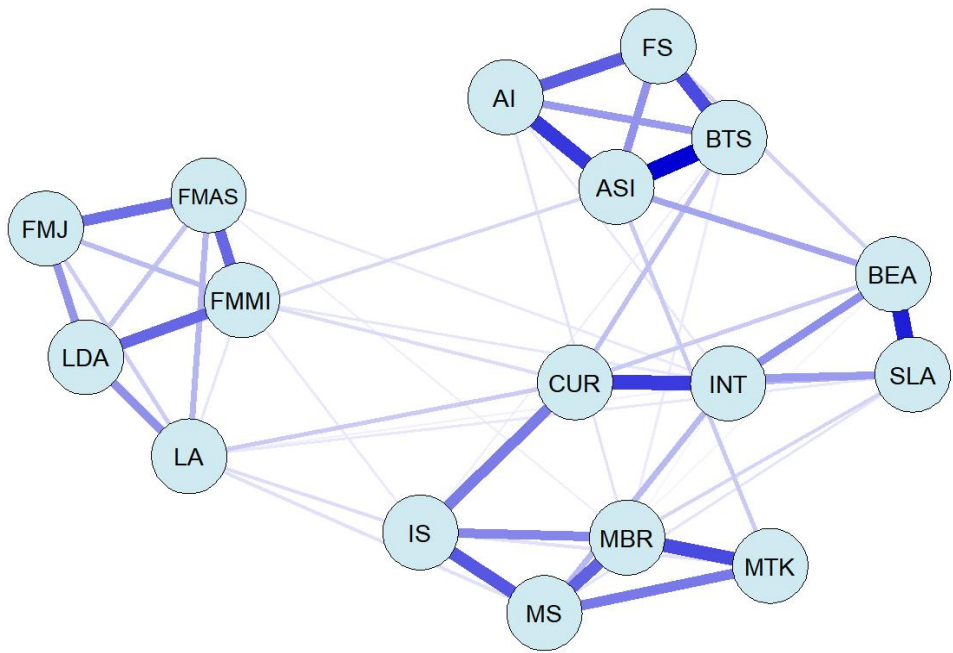

Figure 7. Single Network of childhood adverse experiences

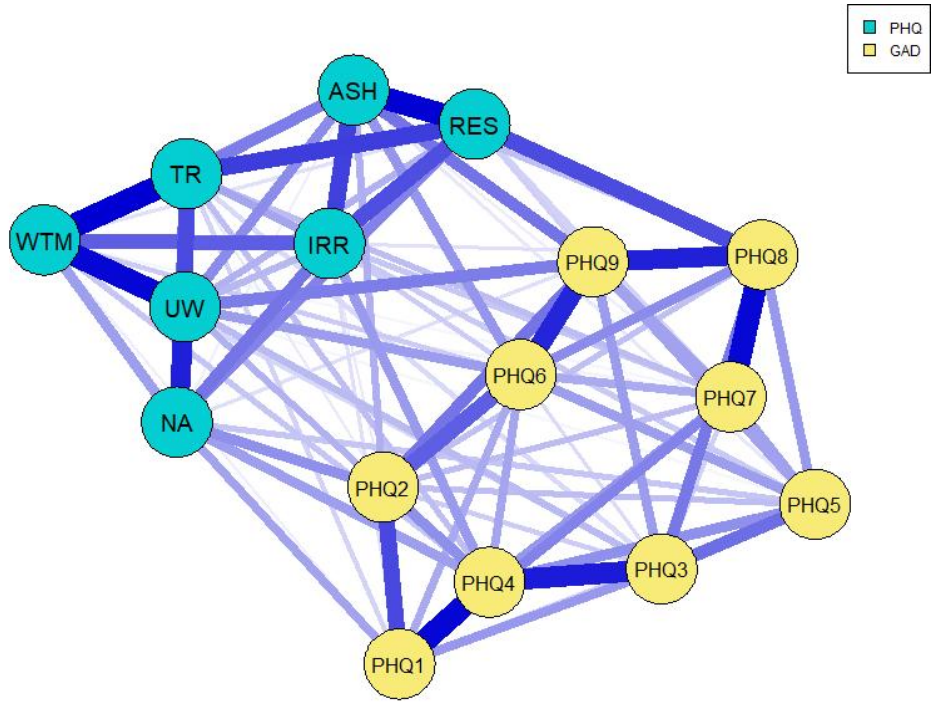

Figure 8. Double Network of depression and anxiety

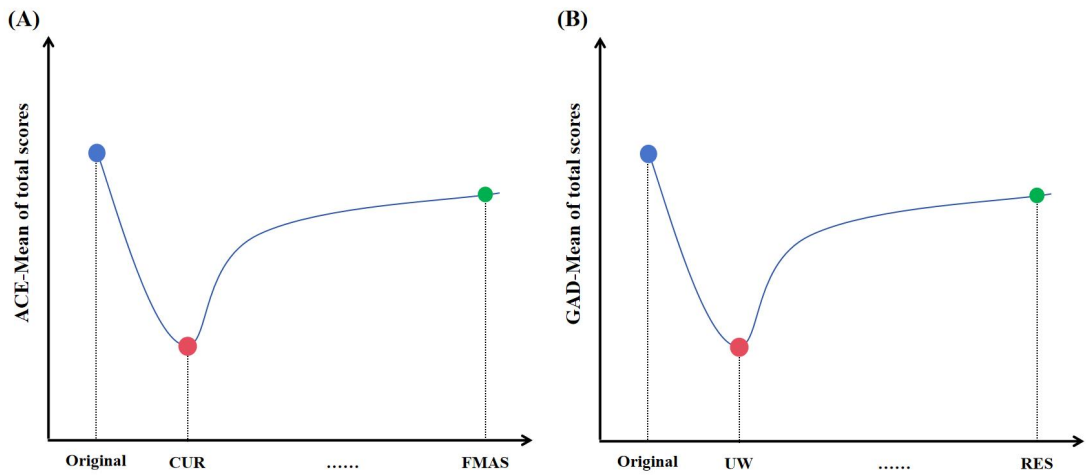

Figure 9. Illustration of Two Types of alleviation Interventions in a Specific Group *i*

for Single and Double Networks

**5 Empirical Application of the Multilevel NIRA**

As an empirical demonstration, we applied the Multilevel NIRA to anxiety data extracted from the 2023 Psychological and Behavioral Investigation of Chinese

Residents (PBICR-2023) to evaluate its effectiveness. The primary goal of our study was to identify the core symptoms that drive the anxiety symptom network among seven specific symptoms. Based on this research objective, we aimed to address the following three questions:

(1) In nested data structures, does the multilevel Ising model exhibit better model performance (as measured by EBIC) compared to the single-level Ising model?

(2) Can Multilevel NIRA identify group-specific core nodes (random intercepts)? And do these group-specific core nodes differ from average core nodes (fixed intercepts)?

(3) Are the average core nodes identified by Multilevel NIRA (fixed intercepts) different from those identified using single-level NIRA?

5.1 Data

The data used in this study were derived from the Psychological and Behavioral Investigation of Chinese Residents 2023 (PBICR-2023). PBICR is a national cross-sectional survey conducted in China aimed at assessing the mental health status and health-related behaviors of Chinese residents (Wu et al., 2024). The PBICR-2023 survey was carried out from June 20 to August 31, 2023, using a multi-stage sampling strategy. The survey directly included four municipalities (Beijing, Tianjin, Shanghai, and Chongqing), the Hong Kong Special Administrative Region, the Macao Special Administrative Region, all 23 provinces, and five autonomous regions of mainland China. The number of cities sampled within each province or autonomous region was determined based on population size, with 2 to 12 cities selected using a random number table. In total, the survey covered 150 cities and 780 communities or villages. Each city recruited at least one investigation team, which conducted face-to-face interviews following systematic training. Respondents participated by accessing an online survey link. Informed consent was obtained prior to data collection. Questionnaire IDs were either entered by the investigators or communicated to participants. For respondents who were cognitively capable but physically unable to complete the questionnaire independently, trained investigators conducted one-on-one interviews and completed the survey on their behalf. After data collection, a two-person back-to-back logic check and data screening process was conducted. A total of 30,054 valid responses were obtained in PBICR-2023.

Since the present study is intended as an applied demonstration of the Multilevel NIRA method, we selected the first 4,000 valid cases for our sample. These 4,000 participants were drawn from 32 regions, including the four municipalities (Beijing, Tianjin, Shanghai, and Chongqing), all 23 provinces, and five autonomous regions of China.

5.2 Measures

5.2.1 Generalized Anxiety Disorder Scale (GAD-7)

The GAD-7 consists of seven items designed to assess the frequency of symptoms associated with generalized anxiety disorder over the past two weeks (Wang et al., 2024). Items are rated on a four-point Likert scale from 0 to 3 (0 = "not at all," 1 = "several days," 2 = "more than half the days," 3 = "nearly every day").

5.2.2 Data analysis

In Step 1, to satisfy the requirements for constructing a multilevel Ising model, we dichotomized the seven GAD-7 items. Following Wang et al. (2025), responses of 1 ("several days"), 2 ("more than half the days"), and 3 ("nearly every day") were recoded as 1 ("present"), while 0 ("not at all") remained coded as 0 ("absent"). In Step 2, because our data are nested (individuals within cities), we applied group mean centering to all predictors prior to modeling. The group mean centering formula is:

$$X_{ij}^c = X_{ij} - \frac{1}{n_j}\sum_{k=1}^{n_j} X_{kj} \qquad (30)$$

Please note that this step of centralization is not done directly. We have constructed a function "IsingFit_multilevel" to integrate the centralization operation with subsequent multilevel Ising model modeling. If centralization is performed before data input into the model, many problems can arise. For example, suppose we are now modeling

$$X_1 \sim X_2 + X_3 + X_4 + X_5 + X_6 + X_7$$

If we initially centered all independent variables, then the dependent variable $X_1$ would also be centered. Generally, in multilevel modeling, the dependent variable is not centered because this would smooth out the effects between groups. Therefore, when constructing functions, we only center the independent variables based on each modeling session.

In Step 3, we employed our custom R function "IsingFit_multilevel" to estimate the multilevel Ising model. From each fitted model we extracted the edge-weight matrix, the fixed intercepts, and the random intercepts. In Step 4, we performed NIRA on both the fixed and random intercepts. For the random-intercept intervention, we focused specifically on the city of Tianjin, simulating ±2-SD shifts in its city-specific intercepts. To facilitate examination of core nodes across all cities, we also provide our "find_Iv_per_level" function, which aggregates NIRA results for each city. Furthermore, we compared the core anxiety-network nodes identified for Tianjin with the average core nodes across all 32 cities in the sample. In Step 5, we fit a single-level Ising model and conducted single-level NIRA on the same dataset, ignoring the nesting by city. We compared (a) the core nodes identified under single-level NIRA with the average core nodes from the multilevel NIRA, and (b) the model fit indices (EBIC) of the single-level versus the multilevel Ising models.

5.3 Result

5.3.1 Comparison of Single-Level and Multilevel Ising Model Outcomes

Figures 10A and 10B display the single-level and multilevel Ising network structures for anxiety, respectively. In these networks, nodes correspond to the specific symptoms assessed by the scale, and edges represent the estimated regression coefficients. Green edges denote positive associations, red edges denote negative associations, and edge thickness reflects the magnitude of the coefficient. Visually, the two networks are highly similar; the primary differences lie in the weights of certain edges—for example, the edge connecting "Nervousness or anxiety" (NA) and

"Restlessness" (RES) appears thicker in the multilevel network than in the single-level network.

Typically, after constructing a network, one would proceed to compute node-centrality metrics and conduct stability analyses. However, because this study's focus is on the development of the Multilevel NIRA algorithm, these steps are omitted here. Instead, we compared model fit indices (EBIC) for the two networks. As shown in Table 1, for every symptom node, the EBIC value of the multilevel Ising model is lower than that of the single-level model, providing empirical support for the appropriateness of multilevel modeling in this context.

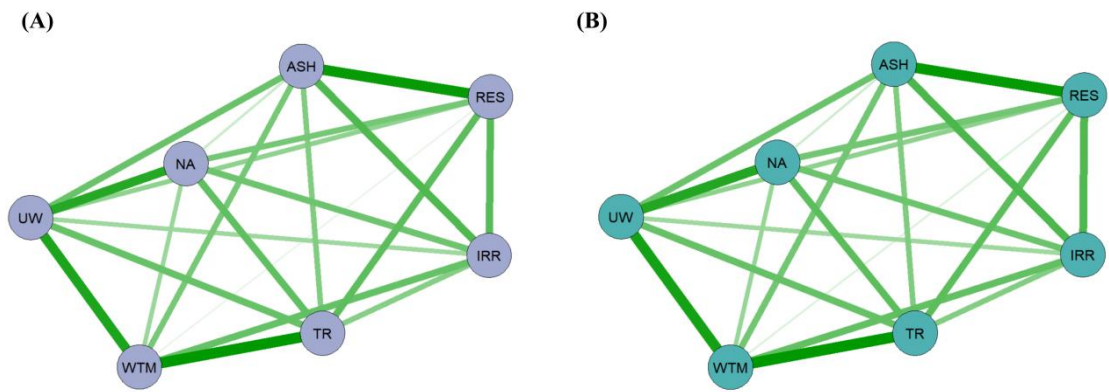

Figure 10. Single-level Ising network(A) and multilevel Ising network(B) of anxiety

Note NA: Nervousness or anxiety, UW: Uncontrollable worry, WTM: Worrying too much, TR: Trouble relaxing, RES: Restlessness, IRR: Irritable, ASH: Afraid something will happen.

Table 1 EBIC values of the two network models

| Node | $EBIC_M$ | $EBIC_S$ |
| --- | --- | --- |
| NA | 3035 | 3048 |
| UW | 2677 | 2681 |
| WTM | 2735 | 2745 |
| TR  | 2546 | 2551 |
| RES | 2900 | 2934 |
| IRR | 2961 | 2978 |
| ASH | 2849 | 2856 |

Note  $EBIC_M$: EBIC value of Multilevel NIRA, $EBIC_S$: EBIC value of Single-level NIRA.

5.3.2 Multilevel NIRA versus Single-Level NIRA Results

To evaluate which symptoms within the anxiety subnetwork play pivotal roles in aggravating or alleviating overall anxiety, we conducted computer-simulated interventions on each node of the GAD-7 network and compared changes in the total network score under both alleviation and aggravation scenarios. In the multilevel model, because both fixed and random intercepts are available and both alleviation and aggravation interventions must be considered, four separate NIRA analyses were performed.

First, when we applied alleviation interventions to the fixed intercepts, reducing the intercept for ASH produced the greatest decrease in average anxiety levels—network scores dropped significantly relative to baseline ($t = 6.21$, $p < 0.001$, p.adjust $< 0.001$)—indicating that ASH is the optimal target for intervention at the population level. Alleviation of RES yielded the second-largest reduction ($t = 5.96$, $p < 0.001$, p.adjust $< 0.001$), whereas alleviating any of the remaining symptoms did not produce significant changes in network score (Figure 11A). Next, aggravation interventions on the fixed intercepts showed that increasing the intercept for ASH led to the largest increase in average anxiety (t = –3.15, p = 0.002, p.adjust = 0.01), again identifying ASH as the prime prevention target; aggravating other symptoms produced no significant score increases (Figure 11B).

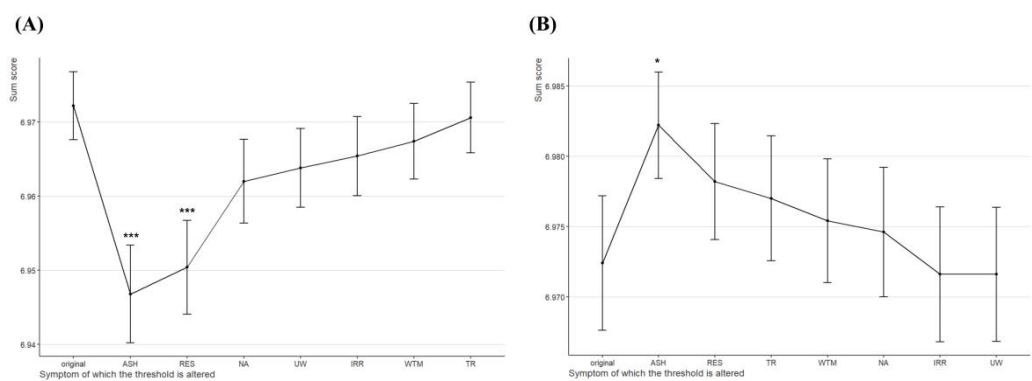

Figure 11. Results of multilevel NIRA fixed intercepts intervention

Second, we focused on the random intercepts for the city of Tianjin. When alleviation interventions were applied to random intercepts, reducing RES resulted in the greatest decline in anxiety for Tianjin residents (t = 11.31, p < 0.001, p.adjust < 0.001), marking RES as Tianjin's most effective local intervention target. Significant reductions in anxiety were also observed when alleviating ASH (t = 10.56, p < 0.001, p.adjust < 0.001), UW (t = 5.45, p < 0.001, p.adjust < 0.001), NA (t = 4.72, p < 0.001, p.adjust < 0.001), TR (t = 2.89, p = 0.003, p.adjust = 0.005), and IRR (t = 2.69, p = 0.007, p.adjust = 0.008) (Figure 12A). We observed that the core symptom identified for Tianjin differs from the average core symptom across all provinces and municipalities, illustrating the effect of the random intercept in enabling group-specific interventions. We then applied aggravation interventions to all nodes, obtaining results similar to those for the fixed intercept. Specifically, for Tianjin residents, increasing the intercept for ASH produced the largest rise in people's anxiety levels, with the total network score significantly higher than the baseline (t = –3.02, p = 0.003, p.adjust = 0.018), indicating ASH as the optimal target for anxiety prevention. However, aggravation of the remaining symptoms yielded non-significant increases in mean anxiety (see Figure 12B).

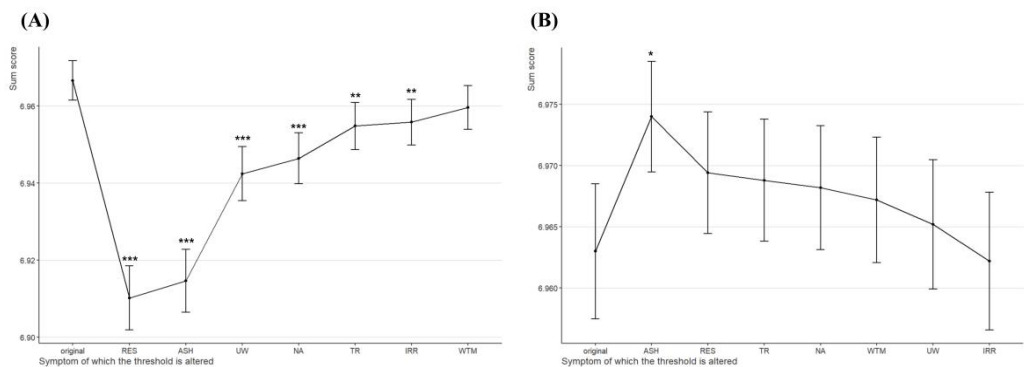

Figure 12. Results of multilevel NIRA random intercepts intervention

In the single-level model—where only one intercept exists—we conducted alleviation and exacerbation interventions. Alleviating WTM yielded the largest reduction in average anxiety (t = 33.53, p < 0.001, p.adjust < 0.001), identifying WTM as the optimal intervention target; significant reductions in anxiety level were also found for TR (t = 33.49, p < 0.001, p.adjust < 0.001), NA (t = 32.31, p < 0.001, p.adjust < 0.001), IRR (t = 31.12, p < 0.001, p.adjust < 0.001), UW (t = 29.93, p < 0.001, p.adjust < 0.001), RES (t = 26.24, p < 0.001, p.adjust < 0.001), and ASH (t = 24.48, p < 0.001, p.adjust < 0.001) (Figure 13A). The differences between single- and multilevel alleviation results underscore the necessity of multilevel modeling for nested data. Aggravation interventions in the single-level model showed that increasing ASH led to the greatest increase in anxiety (t = –32.07, p < 0.001, p.adjust < 0.001); significant increases in anxiety level were also observed for RES (t = –31.05, p < 0.001, p.adjust < 0.001), UW (t = –30.87, p < 0.001, p.adjust < 0.001), IRR (t = –28.35, p < 0.001, p.adjust < 0.001), TR (t = –28.33, p < 0.001, p.adjust < 0.001), WTM (t = –27.00, p < 0.001, p.adjust < 0.001), and NA (t = –23.27, p < 0.001, p.adjust < 0.001) (Figure 13B).

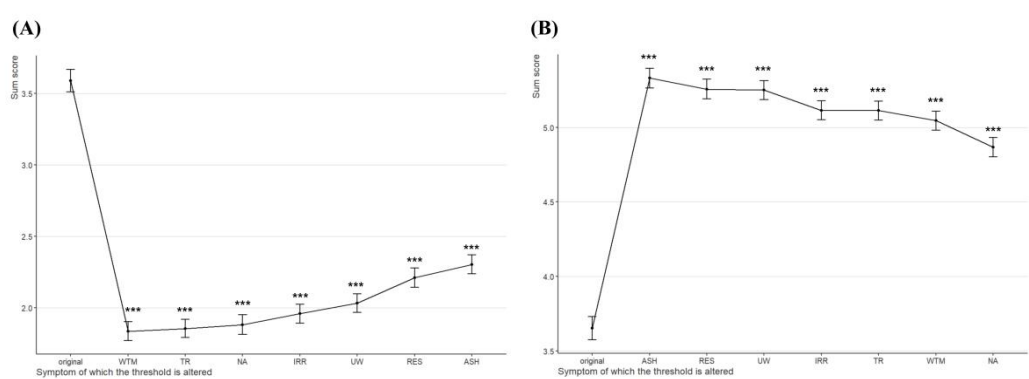

Figure 13. Results of single-level NIRA intervention

## 6 Discussion

In summary, the present study extends the simulated-intervention framework (NIRA) by developing a multilevel NIRA algorithm and provides a comprehensive exposition of its underlying principles—most notably the incorporation of multilevel binary logistic regression and the multilevel Ising model—and detailed modeling procedures. This novel algorithm overcomes NIRA's current inability to handle nested data structures and offers substantial practical utility. By manipulating the spontaneous activation probability of symptoms (i.e., raising or lowering the intercept by two standard deviations), multilevel NIRA simulates in silico symptom exacerbation and alleviation, thereby identifying both optimal prevention and intervention targets. Compared to single-level NIRA, the multilevel extension disentangles between-group and within-group effects. At the between-group level, multilevel NIRA identifies the average optimal target across all groups; at the within-group level, it pinpoints group-specific targets, facilitating precision interventions tailored to each group.

As an empirical demonstration, we applied both single-level and multilevel Ising models to anxiety data from the Psychological and Behavioral Investigation of Chinese Residents 2023 (PBICR-2023). We then implemented single-level NIRA and multilevel NIRA on these respective networks. Across all three NIRA applications, a single node emerged as the optimal target for prevention or intervention, corroborating the view that certain symptoms exert disproportionately large influences on psychopathological processes (Borsboom & Cramer, 2013; Borsboom, 2017). Intriguingly, we observed region-specific differences: in Tianjin, alleviating the "Restlessness" symptom yielded the greatest reduction in network activation, whereas at the national average level, "Afraid something will happen" was identified as the core intervention target. Moreover, multilevel NIRA demonstrated superior performance relative to its single-level counterpart, and the two approaches yielded non-identical core intervention targets— which underscoring the importance of employing multilevel models when analyzing nested data to avoid statistical biases (Wong & Mason, 1985).

Another key advantage of multilevel NIRA is its broad applicability beyond psychopathology. For instance, in aviation research—where understanding how environmental factors and pilot characteristics affect flight performance is critical—prior work has utilized linear mixed models to analyze hierarchical QAR (Quick Access Recorder) data, such as the impact of solar glare on landing distance (one of the indicators used to measure flight performance) (Qiu et al., 2024). However, despite the consideration of the hierarchical nature of QAR data and the use of linear mixed models for modeling, this approach fails to explain causal directionality and cannot comprehensively consider all available indicators, as considering all indicators would require building $N$ models (where $N$ is the number of indicators). Multilevel NIRA effectively addresses this issue. For data nested in flights involving pilots, we can construct a multilevel Ising double network linking pilot psychological factors with flight performance and apply multilevel NIRA. This allows us to consider all indicators comprehensively while also identifying each pilot's unique core psychological factors.

In the process of developing our multilevel NIRA algorithm, we identified a particularly intriguing element, namely the inverse temperature coefficient $\beta$ mentioned above. Over the past decade, most psychological studies that employ cross-sectional data for network analysis have implicitly fixed the inverse temperature coefficient at 1, since researchers have been primarily concerned with the pattern of influences among variables rather than with the broader structure of network symptoms. However, the concept of network temperature is itself a valuable metric, as it quantifies the stability and predictability of the network system (Scheffer et al.,

2024). Here, "temperature" denotes the degree of randomness in the possible states that nodes may assume (for instance, 0 or 1 in the Ising model), and it relates to the inverse temperature coefficient as follows:

$$\beta = \frac{1}{k_\mathrm{B} T} \qquad (31)$$

Where $\beta$ is the inverse temperature coefficient, $T$ is the temperature, and $k_\mathrm{B}$ is the Boltzmann constant.

Moreover, temperature is frequently linked to entropy: the higher the temperature of a network, the greater its entropy, the less stable the network becomes, and the more randomly nodes take on their values. Empirical investigation of network temperature began with the depressive-symptom network study conducted by Grimes et al. (2025). In that longitudinal work, adolescent depressive networks were estimated at each age from 11 to 14 years using cross-sectional Ising models; the authors observed a progressive decline in network temperature with increasing age, suggesting that depressive symptoms become more stable as adolescents grow older. A limitation of their approach, however, is that by estimating a separate Ising model at each age, they potentially overlooked the autoregressive effects of individual symptoms. While such omissions may have minimal impact in designs with only a few measurement waves, they become non-negligible in intensive longitudinal studies (Lischetzke, 2024).

At present, our multilevel NIRA algorithm is optimized for nested cross-sectional data and does not explicitly model temporal effects. Future extensions of this algorithm could incorporate time dynamics and further integrate the concept of network temperature, thereby enabling us to examine how symptom stability evolves over time.

As with NIRA, multilevel NIRA also exhibits certain limitations. First, the current implementation of multilevel NIRA is framed entirely within a multilevel Ising model context, which mandates that all network predictors be binary (0/1) variables. This requirement substantially restricts the algorithm's applicability. Although researchers can achieve binarization through transformation, this may lead to psychometric problems caused by changes in the response patterns of research tools. Because psychological scale data are ordinarily treated as continuous during analysis, future work could, following the logic of our present study, extend NIRA to operate within a Gaussian Graphical Model (GGM) or a Mixed Graphical Model (MGM) framework to better accommodate continuous—and mixed—data types.

Second, it remains unclear whether the intervention effects simulated by multilevel NIRA correspond to effects observed in genuine manipulations, suggesting that generalizations from multilevel NIRA results should be made with caution. Simulated intervention estimates derived from multilevel NIRA cannot fully substitute for empirical intervention outcomes, as real-world interventions are seldom as targeted or idealized as those modeled in simulation. In practice, an intervention may inadvertently affect multiple symptoms rather than a single, isolated symptom. For example, a common cold medication might simultaneously reduce both "runny nose" and "sore throat" symptoms, thereby complicating the alignment between simulated and actual intervention effects. Nonetheless, multilevel NIRA offers unique advantages in research scenarios where ethical or practical constraints preclude the execution of randomized experimental studies.

**7 Conclusion**

In this study, we built upon the original NIRA framework to develop a multilevel NIRA algorithm and have provided a detailed exposition of its underlying modeling principles and procedural steps. Our algorithm overcomes the prior limitation of NIRA in handling cross-sectional nested data and substantially broadens its practical applicability: it not only simulates causal processes in nested network analysis but also allows for tailored simulated interventions at each stratum of the Level-2 variable. We very much hope that this contribution will prove valuable to researchers in need of such methods.